\documentclass[usegraphicx,useAMS]{mn2e}

\newcommand\heii{\hbox{He\,{\sc ii}}}
\newcommand\hi{\hbox{H\,{\sc i}}}
\newcommand\teff{\mbox{$T_{\rm eff}$}}

\def\eps@scaling{1.0}%
\newcommand\epsscale[1]{\gdef\eps@scaling{#1}}%

\newcommand\plotone[1]{%
 \centering
 \leavevmode
 \includegraphics[width={\eps@scaling\columnwidth}]{#1}%
}%

\newcommand\plottwo[2]{%
 \centering
 \leavevmode
 \columnwidth=.48\textwidth
 \includegraphics[width={\eps@scaling\columnwidth}]{#1}%
 \hfil
 \includegraphics[width={\eps@scaling\columnwidth}]{#2}%
}%

\title{A-star envelopes: a test of local and non-local models of convection}

\author[Kupka \& Montgomery]
       {F. Kupka$^{1,2}$ \& M. H. Montgomery$^3$ \\
       $^{1}$Institut f\"ur Astronomie, Universit\"at Wien,
       T\"urkenschanzstra\ss e 17, A-1180 Wien, Austria \\
       $^{2}$Institut f\"ur Mathematik, Universit\"at Wien,
       Strudlhofgasse 4, A-1090 Wien, Austria \\
       $^3$Institute of Astronomy, University of Cambridge, Madingley Road,
       Cambridge CB3 0HA, United Kingdom}

\date{\vspace*{-3em} Accepted 2001? December 15.}

\pagerange{\pageref{firstpage}--\pageref{lastpage}}
\pubyear{2001}

\begin{document}

\maketitle

\label{firstpage}

\begin{abstract}

We present results of a fully non-local, compressible model of
convection for A-star envelopes. This model quite naturally reproduces
a variety of results from observations and numerical simulations which
local models based on a mixing length do not. Our
principal results, which are for models with \teff\ between 7200~K and
8500~K, are the following: First, the photospheric velocities and
filling factors are in qualitative agreement with those derived from
observations of line profiles of A-type stars. Second, the \heii\ and
\hi\ convection zones are separated in terms of convective flux and
thermal interaction, but joined in terms of the convective velocity
field, in agreement with numerical simulations. In addition, we attempt
to quantify the amount of overshooting in our models at the base of the
\heii\ convection zone.

\end{abstract}


\section{Introduction}

Over the last five decades the most frequently used approach to
describe stellar convection has been the mixing length theory (MLT,
Biermann 1948, B\"ohm-Vitense 1958). However, the great simplicity
achieved by describing convection in terms of local variables is only
attained at the cost of trade-offs, the most important of which is the
specification of a mixing length that can neither be derived from
rigorous theory nor from observations. More recently, turbulence
models, e.g.\ by Canuto et al.\ (1996, hereafter the CGM model) have
been used to improve the MLT expressions. These convection models still
provide a local expression for the temperature gradient and contain the
specification of a scale length $l$. The latter also holds for non-local
versions of the MLT which were proposed to account for convective
overshooting. However, the intrinsic non-locality of this problem has
prohibited a satisfactory solution within the context of models that
use any form of local scale length (see Renzini 1987 and Canuto 1993).

This difficulty is naturally avoided by numerical simulations which
have come into use during the last decade as a tool to study stellar
surface convection. Simulations in 3D have mostly been devoted to solar
convection (Nordlund \& Dravins 1990, Atroshchenko \& Gadun 1994, Kim
\& Chan 1998, Stein \& Nordlund 1998), while 2D simulations have been
used for more extended computations over the HR diagram (cf.\ Freytag
1995 and Freytag et al.\ 1996). Such calculations can include
the entire convective part of a stellar envelope only for the case of
A-stars (and some types of white dwarfs). Even then, the computational
efforts become considerable, especially when realistic microphysics is
used and thermally relaxed solutions are required. To use simulations for
complete stellar models is thus beyond the range of present computer
capabilities (cf.\ Kupka 2001).

Another alternative was pioneered by Xiong (1978) who used the Reynolds
stress approach. This approach had previously been applied in atmospheric as well
as in engineering sciences. But even in its most recent version (Xiong et
al.\ 1997) his formalism still uses a mixing length to calculate the
dissipation rate $\epsilon$ of turbulent
kinetic energy. Canuto (1992, 1993) and Canuto \& Dubovikov (1998, hereafter
CD98) abandoned the use of a mixing length in their Reynolds stress models.
These models provide both the mean quantities of stellar structure (temperature $T$,
pressure $P$, luminosity $L$, and mass $M$ or radius $r$) as well as
the second order moments (SOMs) of temperature and velocity fields created
by stellar convection (turbulent kinetic energy $\rho K$, temperature
fluctuations $\overline{\theta^2}$, convective flux $F_{\rm C}=
c_p \rho \overline{w\theta}$, vertical turbulent kinetic energy
$\frac{1}{2} \rho \overline{w^2}$, and the dissipation rate $\epsilon$) as the solution of
coupled, non-linear differential equations. Their models are thus fully
non-local on the level of second order moments. Numerical solutions of
these models for the case of idealized microphysics have been presented
by Kupka (1999a) and Kupka (1999b, 2001). The same equations, using
realistic microphysics, were later
solved for the \heii\ convection zone of A-stars (Kupka \&
Montgomery 2001; these results were first discussed in Canuto 2000).

In this paper, we present solutions for complete A-star envelopes.
Numerically, this problem is easier than that of convection in the Sun
since A-stars are hotter and therefore have thinner convection zones.
In addition, A-stars reveal the shortcomings of local convection models
more clearly, as their less efficient convection is much more sensitive
to details in the modelling. Depending, for example, on whether an $\alpha$
of 0.5 or 1.5 is chosen in MLT for a main sequence star with \teff\
$\sim 7500$~K, an envelope may either have a mostly radiative temperature
gradient or still contain a nearly adiabatic region. This holds for any
of the convection models which rely on a convective scale length. Hence,
the efficiency of convection in the envelopes of A-stars has remained an
open problem and makes them a logical as well as a promising starting point 
for our study.

In the following, we give an outline of the physics and the numerical
procedure used to compute our envelope models (the discussion of the
moment equation formalism is self-contained, so that readers unfamiliar
with it can skip ahead without difficulties). Results are then presented
for a sequence of models which differ from each other only in \teff. 
We include a model with lower gravity in order to illustrate the effect of
a change in $\log g$. Finally, we show that the non-local convection
model agrees with the known observational constraints and the results
of numerical simulations, whereas local models are fundamentally
unable to do this.

\section{Description of Model}   \label{Sect2}

\begin{figure*}
\plottwo{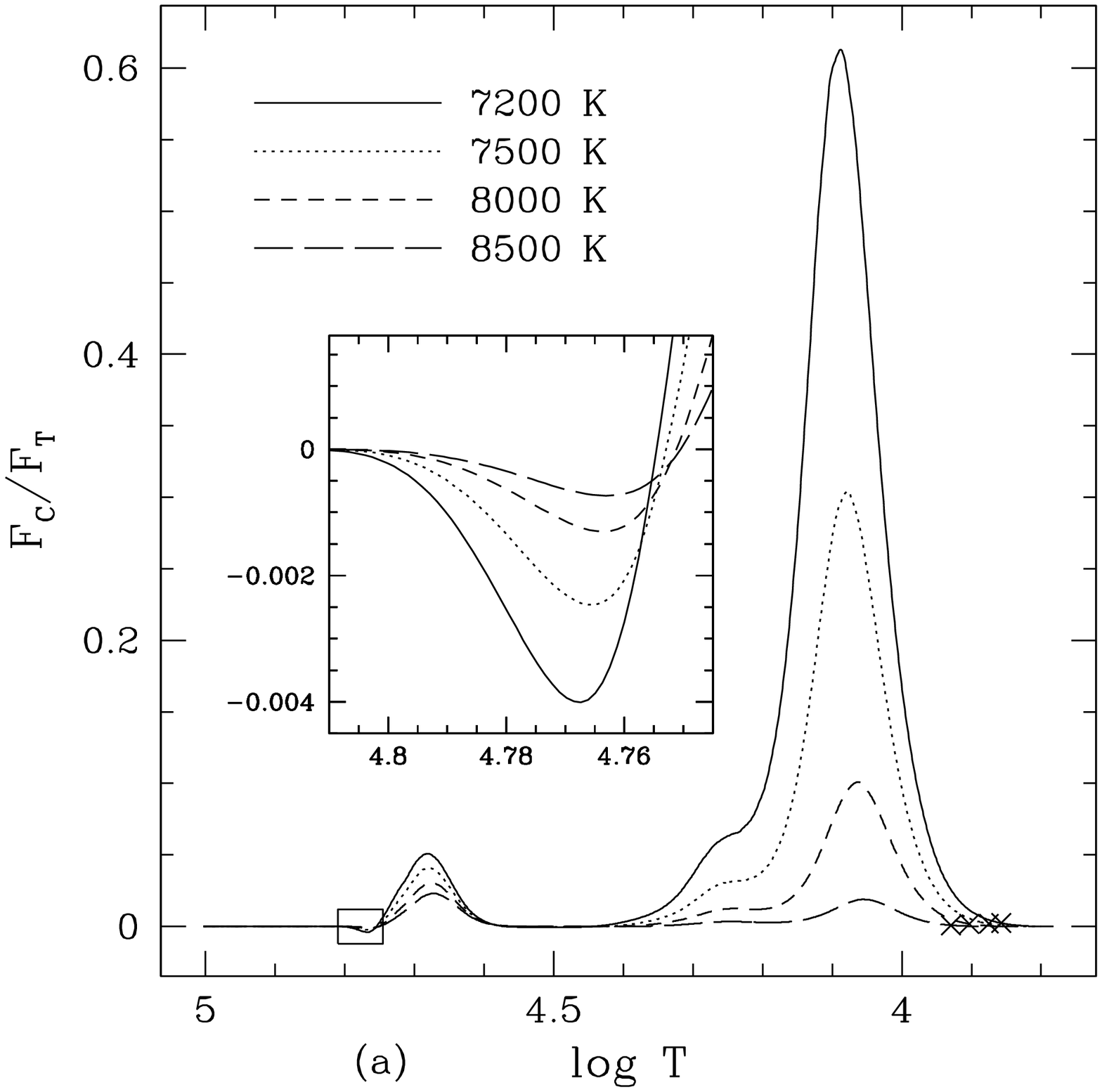}{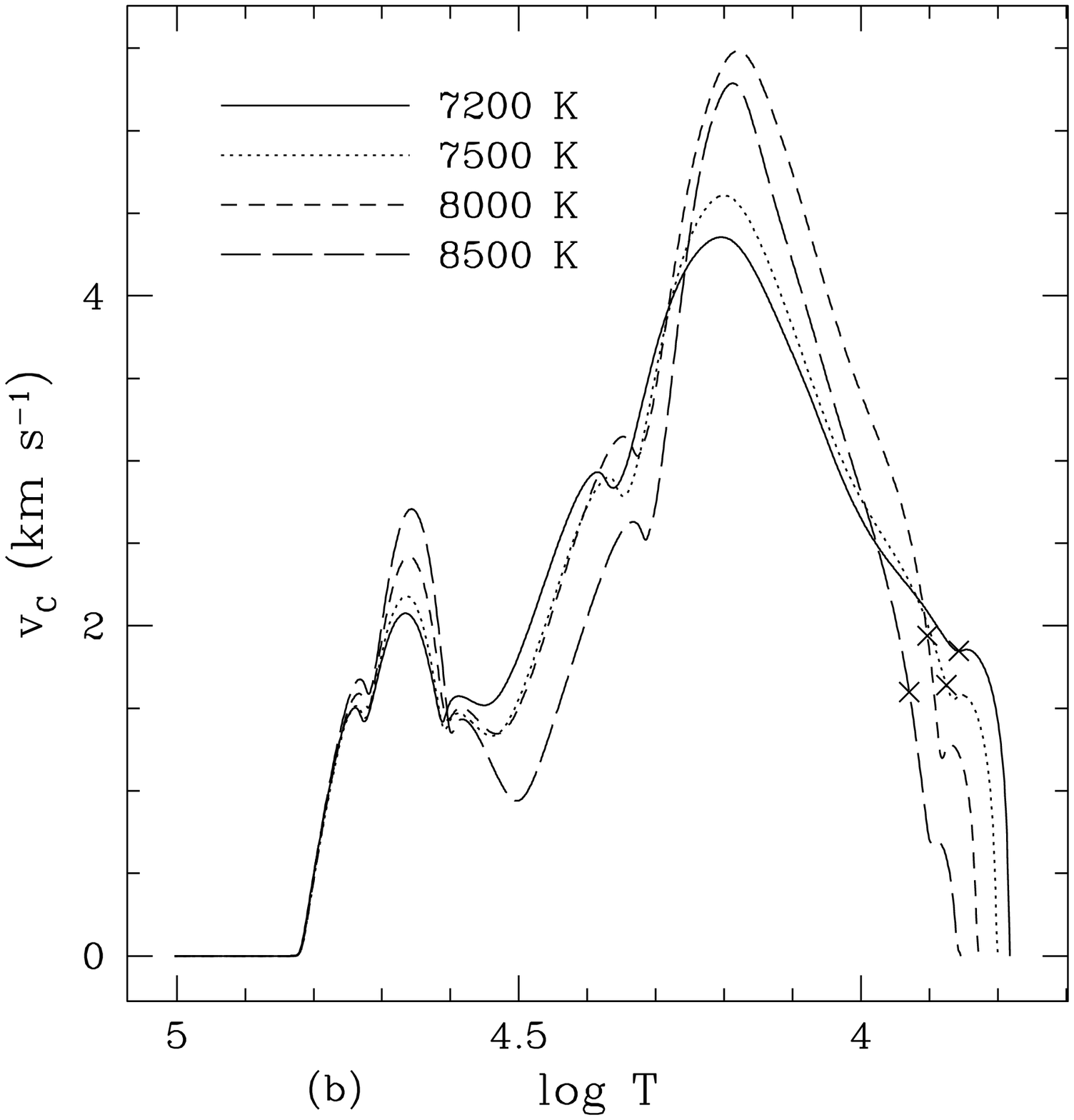}
\caption{(a) The fraction of the flux carried by convection for four models
with the indicated effective temperatures, where we have taken $\log T$
as our radial variable; $\log g = 4.4$ and $Z = 0.02$ for all the models. The cross on
each curve near $\log T \sim 3.9$ shows the location where $\tau=2/3$ for
each model, and the inset is an enlargement of the indicated overshooting region. Convection becomes more dominant with decreasing
$T_{\rm eff}$, but the two convection zones remain well-separated
in terms of the convective flux. This holds also for
$\overline{\theta^2}$ (cf.\ Kupka 2002).
(b) The same as (a) but for the rms convective velocities. In contrast,
we see that the convection zones {\em are} connected in terms of the
velocity field (and also in terms of $K$ and $\epsilon$, see Kupka
2002). Thus, the two zones can be thought of as being separate
thermally but not dynamically. In addition, we see that the
photospheric velocities all lie in the range 1.5--2 km~s$^{-1}$, in
agreement with the lower limit of derived micro- and macroturbulence
parameters (Varenne \& Monier 1999, Landstreet 1998).
\label{models}
}
\end{figure*}

The convection model used here is an extension of the CD98 model which
requires the solution of five differential equations of first order in
time and second order in space for $K$, $\overline{\theta^2}$,
$J=\overline{w\theta}$~$=F_C/(\rho c_p)$, $\overline{w^2}$, and
$\epsilon$, and of an additional equation for the time evolution of $T$
(cf.\ equations (1)--(5) and (8) in Kupka 1999b). This system is
completed by an equation for the total pressure (``hydrostatic
equilibrium'' including turbulent pressure, equation (7) in Kupka 1999b)
and for the mass (``conservation of mass'').  We solve this set of
differential equations on an unequally spaced mass grid, with the zoning
chosen so as to resolve the gradients in the various quantities.

Compared to the model discussed in Kupka (1999b) the following changes
and extensions have been included:  a) instead of using high Peclet
number limits we apply the full form of the CD98 model for the SOMs. We
thus take advantage of a better theoretical underpinning of the
influence of radiative loss rates on two time scales in the equations
for $\overline{\theta^2}$ and $J$, $\tau_{\theta}$ and
$\tau_{p\theta}$, which are provided by a well-tested turbulence model
(see CD98 for a summary).  b) The Prandtl number is set to $10^{-9}$ as
a typical value for the outer part of A-star envelopes (values up to 2
orders of magnitude larger than this do not alter our results). c) With
the exception of the pressure correlations $\overline{p'w}$ and
$\overline{p'\theta}$ which require further study (see Kupka \& Muthsam
2002), the complete form of the ``compressibility terms'', equations
(42)--(48) of Canuto (1993), is used to extend the CD98 model to the
non-Boussinesq case. Hence, we now also include the effect of a
non-zero gradient in the turbulent pressure $p_{\rm turb}$ on the
superadiabatic gradient $\beta$. d) We use a more advanced model for
the third order moments (TOMs) published in Canuto et al.\ (2001), although
with a different form for the fourth order moments (see
Kupka 2002). If, instead, the original form for the 
fourth order moments is used, the models with $\teff \geqslant 8000$~K,
discussed in Figure~\ref{models}, show less efficient convection, with
the opposite being true for the cooler models. In both cases, however,
the results are qualitatively the same as the results we present here.
As in Kupka \& Montgomery (2001), we use a relation similar to equation
(37f) in Canuto (1992) and thus avoid a downgradient approximation
for the flux of $\epsilon$ (such as equation (6) of Kupka 1999b).
e) The effect of stratification on the pressure correlation
time scales, $\tau_{pv}$ and $\tau_{p\theta}$, was accounted for
following Canuto et al.\ (1994). Likewise, the time scales
$\tau_{\theta}$ and $\tau_{p\theta}$ include a correction for the
optically thin regime of stellar photospheres (cf.\ Spiegel 1957),
while for consistency, the expression for the radiative flux $F_{\rm
r}$ was taken from the stellar structure code we use for our initial
models and boundary conditions (see Pamyatnykh 1999, it assumes the
diffusion approximation for $\tau\geqslant 2/3$, but differs from it by
a ``dilution factor'' for optical depths $\tau < 2/3$).

More details on these alterations and comparisons with numerical simulations
are discussed in Kupka (2002) and in Kupka \& Muthsam (2002). With one
exception we have used the original constants of Canuto (1993), CD98, and
Canuto et al.\ (2001). We consider their adjustment to be of little use, because
in case of failure it is usually the entire shape of the functional relation
which is at variance with measurements or simulations (cf.\ the MLT example
in Sect.~\ref{Sect4}). The one exception we have made is the high efficiency
limit of $\tau_{p\theta}$, for which the CD98 model appears to predict values
too low in comparison with simulations for idealized microphysics (see Kupka
2001), and also in comparison with a previous model (Canuto 1993). Most likely
this is due to an isotropy assumption in its derivation and we thus use
a $\tau_{p\theta}$ increased by a factor of 3 as suggested in Kupka
(2001). This problem will be thoroughly discussed in Kupka \& Muthsam (2002).

A numerical approach to solve the resulting system of equations was
briefly described in Kupka (1999a,b); a comprehensive discussion of the
code will be given in Kupka (2002). Here we only outline the solution
procedure from the viewpoint of stellar structure modelling. We start
from an envelope model computed with the code described in Pamyatnykh
(1999), where the equation of state and opacity data are from the OPAL
project (Rogers et al.\ 1996, Iglesias \& and Rogers 1996). The
metallicity, \teff, surface $\log g$, and total stellar radius
$R_{\star}$ are taken from this model and held constant during
relaxation. We place some 200 mass shells from the mid photosphere
(with $\tau_{\rm ross}\sim 10^{-3}$) down to well below the
\heii\ convection zone. Having embedded the convection zones within
stably stratified layers, we can use the boundary conditions of Kupka
(1999b) for the SOMs (cf.\ Kupka 2002). For the mean structure
quantities we keep $r$, $T$, and $P$ fixed to their values at the upper
photosphere of the input model, while a constant $L$ is enforced at the
bottom. The complete system is integrated in time (currently by a
semi-implicit method) until a stationary, thermally relaxed state is
found. The mass shells can be rezoned to a different relative size to
resolve, e.g., steep temperature gradients that may appear
and/or disappear during convergence. The radiative envelope below the
convection zones may then be obtained from a simple downward
integration.

Generating a complete stellar model would require fitting such an envelope
onto a stellar core, which in turn requires iterating the envelope parameters
(\teff , $\log g$, $R_{\star}$) to achieve a match of $P$, $T$, and $r$
at the core/envelope interface. Since we have not yet computed evolutionary
models, we have not needed to do this, although this would be a straightforward
extension of our work.

\section{Results}   \label{Sect3}

Figure~\ref{models} shows the central results of this paper: both the
\heii\ and \hi\ convection zones appear quite separate when the
quantity which is examined is the convective flux (a), but completely
merged in terms of the convective velocity field (b). Thus, to obtain a
self-consistent solution, one must solve the equations for the entire
region simultaneously.

From Figure~\ref{models}a (and Figure~\ref{fkin}), we see that the mid
to the upper photospheres of these models (the crosses indicate the point
where $\tau=2/3$) are essentially radiative, as they are in the local CGM and
MLT models. Thus, the temperature and density structure of both the local
and non-local models are virtually identical at small optical depths, which
justifies our use of the local models as an outer boundary condition
for the non-local models.

In Table~\ref{params}, we list these results.  Since the \heii\ and \hi\
convection zones are well-separated in terms of $F_{\rm C}/F_{\rm T}$,
we have listed their maximum fluxes separately (columns 3 and 4). For the
convective velocity, $v_{\rm C}=(\overline{w^2})^{0.5}$, we have listed
just a single maximum since this quantity is large throughout the entire
region (column~6); the same holds for the relative turbulent pressure
(column~8). Since all of these maxima occur below the stellar surface,
we have also listed the photospheric ($\tau=2/3$) values of $v_{\rm C}$
and $p_{\rm turb}/p_{\rm tot}$ (columns~7 and~9).

Finally, in Figure~\ref{fkin} we plot the kinetic energy flux as a
function of $\log T$, for the four different models from
Figure~\ref{models}. Besides the fact that the cooler models have
larger fluxes, which is to be expected, we see from the magnitudes of
these fluxes that $F_{\rm kin}$ is essentially negligible for the
models we have examined. We are thus in a different regime from that of
the Sun, where $|F_{\rm kin}/F_{\rm T}|$ may be as large as 20 per cent
(cf.\ Stein \& Nordlund 1998, Kim \& Chan 1998).

In addition to these results, we have also run low- and
high-metallicity models ($Z=0.006, 0.06$, respectively). We find that
for the low-$Z$ models, $(v_{\rm C})_{\rm max}$ decreases $\la 3$ per cent
while $(F_{\rm C})_{\rm max}$ increases by $\la 10$ per cent, with the
opposite trends for the high-$Z$ models. While these changes are not
large, we note that they would be enhanced by the use of non-grey
atmospheres. On the other hand, reducing $\log g$ (to a value still
consistent with a main sequence object)
results in much weaker convection caused by a lower density and hence
smaller heat capacity of the fluid, as shown by the last model in
Table~\ref{params}, which is taken from an (MLT based) evolutionary sequence
of a 2.1~$M_\odot$ star. 

\section{Discussion}   \label{Sect4}

\begin{table*}
\vspace*{-0.1cm}
 \centering
 \begin{minipage}{140mm}
  \caption{Convection zone parameters obtained with the non-local
           model. The overshooting (OV) is measured from the
           minimum of $F_{\rm C}/F_{\rm T}$ (shown in the inset of 
	   Figure~\protect\ref{models}a) to the point where $|F_{\rm C}/F_{\rm T}|\sim 10^{-6}$.}
  \begin{tabular}{clccccccc}
 $T_{\rm eff}$ & $\log g$ & \multicolumn{2}{c}{$(F_{\rm C}/F_{\rm T})_{\rm max}$}
 & OV & $(v_{\rm C})_{\rm max}$ & $(v_{\rm C})_{\tau=2/3}$ &
        $(p_{\rm turb}/p_{\rm tot})_{\rm max}$ &
        $(p_{\rm turb}/p_{\rm tot})_{\tau=2/3}$ \\
 (K) &  & \heii & \hi\ & (in $H_p$) & (km~s$^{-1}$) & (km~s$^{-1}$) & &
  \\
 8500 & 4.4 & 0.023 & 0.019 & 0.44 & 5.29 & 1.60 & 0.131 & 0.043 \\
 8000 & 4.4 & 0.030 & 0.100 & 0.46 & 5.48 & 1.94 & 0.146 & 0.068 \\
 7500 & 4.4 & 0.041 & 0.303 & 0.45 & 4.61 & 1.64 & 0.105 & 0.053 \\
 7200 & 4.4 & 0.051 & 0.612 & 0.46 & 4.36 & 1.85 & 0.100 & 0.069 \\
 6980 & 3.53 & 0.038 & 0.164 & 0.52 & 5.33 & 1.40 & 0.130 & 0.042 \\
\label{params}
\end{tabular}
\end{minipage}
\vspace*{-0.65cm}
\end{table*}

\begin{figure}
\plotone{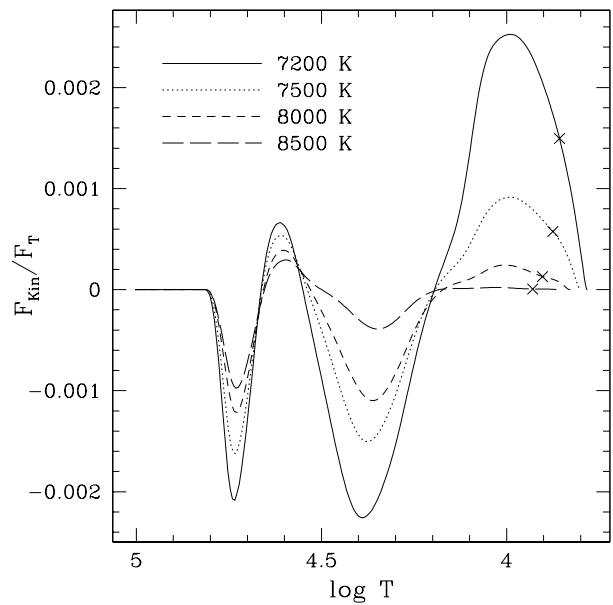}
\caption{The kinetic energy flux as a function of $\log T$, for the four
different models from Figure~\ref{models}. As expected, the cooler models
have larger fluxes. Most significantly, however, these numbers show that
$|F_{\rm kin}|$ is essentially negligible for the models we have
examined, in contrast to the case of the Sun, where it may be as large
as 20 per cent (cf.\ Stein \& Nordlund 1998, Kim \& Chan 1998).
\label{fkin}
}
\end{figure}

The fact that $F_{\rm kin}$ is positive in the photosphere for each of
these models (Figure~\ref{fkin}) means that the skewness of spectral
lines produced in this region is also positive, and that the
corresponding filling factors for rising versus falling fluid elements
is less than 1/2 (cf.\ CD98). This is in agreement with the
observations of line profiles in A-stars (Landstreet 1998). In
the future, quantitative comparisons with such observational data will
provide some of the most stringent tests of this model.

\begin{figure}
\plotone{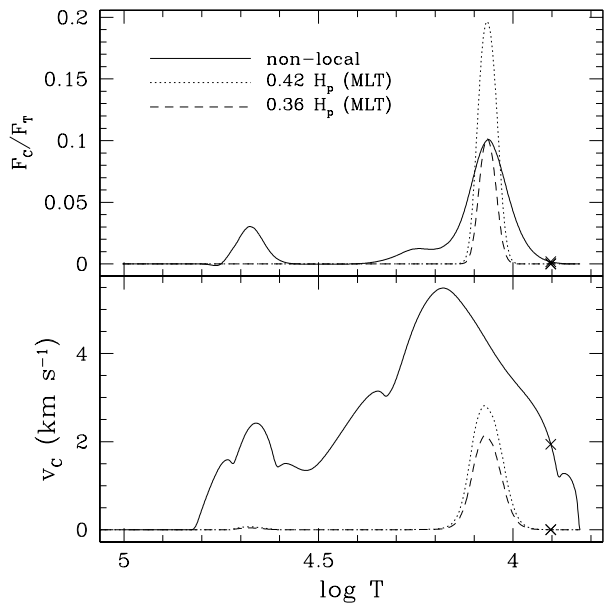}
\caption{A comparison of the convective fluxes (upper panel) and 
velocities (lower panel) for the non-local model and for two MLT models
($l = 0.36$ and $0.42 H_p$), for
$T_{\rm eff}=8000$~K. We see that the MLT models are unable to match
simultaneously the flux in the \hi\ and \heii\ convection zones, or
even to match both the maximum flux and the width of just the \hi\
convection zone. In addition, the photospheric velocities of the MLT
models are $\sim 3$ orders of magnitude smaller than those of the
non-local model.
\label{mltcomp}
}
\end{figure}

As previously mentioned, the \heii\ and \hi\ convection zones may be
thought of as being thermally disconnected but dynamically coupled, a
situation which is impossible within the context of MLT (or CGM) models.
A further shortcoming of MLT is shown in Figure~\ref{mltcomp}. The convective
flux of two MLT models, with mixing lengths of $l = 0.36$ and $0.42 H_p$,
respectively, is plotted along with the flux from the non-local solution
(upper panel).  First, we see that it is impossible for the MLT models to
match simultaneously the flux in both the \heii\ and \hi\ convection
zones (at least with the same mixing length). Second, even if we try to
model only the \hi\ convection zone, fixing the mixing length so as to
match the maximum flux results in a convection zone which is much too
narrow. In addition, this produces photospheric velocities which are $\sim 3$
orders of magnitude smaller than those of the non-local model and the
observations (lower panel, Figure~\ref{mltcomp}, see also Sect.~\ref{Sect5}).
We note that since the upper photosphere is optically thin and
therefore locally stable against convection, local convection models
will always predict convective fluxes which are extremely small (or
zero), even for values of $\alpha$ which are ``unreasonably'' large.

As a further test of our results we have compared them with 2D
simulations by Freytag (1995), Freytag et al.\ (1996), and additional
models provided by Freytag (2001, private communication).  We find
agreement with the following results from our calculations:  a) Models
over the entire range of A-type main sequence stars with \teff\ up to
8500~K have their \hi\ and \heii\ convection zones dynamically
connected. The vertical mean velocities in the overshoot regions around
$\log T \sim 4.4$ are of order 1.5 to 3 km~s$^{-1}$. b) There is
considerable overshooting (OV) below the \heii\ convection zone.
However, the 2D simulations yield a size of the OV region which is 3
times larger in terms of radius and also much larger in terms of $H_p$,
for the entire range of models in Figure~\ref{models} and
Table~\ref{params}. Such differences are anticipated from a comparison
of numerical simulations in 2D and 3D (Muthsam et al.\ 1995, see also
Fig.~1 in Kupka 2001). More detailed examples demonstrating that 2D
simulations yield upper limits for the (3D) OV extent will
be given in Kupka \& Muthsam (2002). c) The maximum of $F_{\rm C}$
and the temperature gradient in the \hi\ convection zone for the models
with $\teff \geqslant 8000$~K are in good agreement with those of
the 2D simulations. However, for the models with lower \teff, the 2D
simulations yield higher convective fluxes and lower temperature gradients
and, hence, the two convection zones merge thermally at a \teff\ which is
$\sim 200$~K to $\sim 300$~K higher than in our non-local models.

Apart from the differences between 2D and 3D convection, one important
reason for discrepancies is the effect of ionization (cf.\ also
Kupka 2002). Briefly summarized, the current convection model
assumes an ideal gas equation of state for the purpose of computing the
ensemble averages in the expression for the convective flux. Using an
improved, although approximate, expression for the convective (enthalpy)
flux, we estimate that this assumption introduces errors of order 15--20
per cent in the convective flux.
Finally, a potentially significant source of discrepancies between our
models and the 2D simulations is the use of a different equation of state
and opacities (OPAL, Rogers et al.\ 1996 vs. ATLAS6, Kurucz 1979) and the non-diffusive law we use for
the photospheric radiative flux (see Sect.~\ref{Sect2}). This does
not allow us to make a detailed quantitative comparison of model sequences.
Thus, we have had to restrict ourselves to only a qualitative discussion.

\section{Conclusions}  \label{Sect5}

Using a fully non-local, compressible convection model together with a
realistic equation of state and opacities, we have calculated envelope
models for stellar parameters appropriate for A-stars. In examining the
results of this model, we have found many points of agreement both with
observations and with numerical simulations.

First, our photospheric velocities are consistent with the lower limit
of the typical micro- and macroturbulence parameters found for A-stars
(1.5--2 km s$^{-1}$, see Varenne \& Monier 1999 and Landstreet 1998).
Line blanketing should further increase these values. We expect a smoother
$v_{\rm C}(r)$ (without small minima as in Figures~\ref{models}b and
\ref{mltcomp}) from an improved treatment of fourth order moments and
inclusion of $\overline{p'w}$ (cf.\ Sect.~\ref{Sect2}). Second, we find
that the filling factor for rising fluid elements in the photospheres of
our models is less than 1/2, also in agreement with observations of line
profiles in A-stars. Third, we find
in the temperature range 7200~K to 8500~K that the \heii\ and \hi\ zones
are well-separated in terms of the convective flux but {\em not} in terms
of the convective velocity field. The two zones are thus in some sense
thermally separated but dynamically joined. This feature is also shown
by the numerical simulations. Finally, we find an OV at the
base of the \heii\ convection zone of $\sim 0.45 H_p$. The numerical
simulations find an even larger OV, but this may also be due to the fact
that they were done in 2D. We note that in all cases we find a nearly
radiative temperature gradient in the OV region, whereas the velocities
in this region remain quite large, within an order of magnitude of their
maxima within the convection zone
($\sim 0.5$ km s$^{-1}$).

In addition, the non-local model yields smaller temperature gradients
than the local model of Canuto et al.\ (CGM, 1996). Such a comparison
with MLT is more difficult due to the large range of
$\alpha$ in current use. Nevertheless, we have found evidence that for
main sequence models $\alpha$ has to be decreased from values of $\sim
1.0$ at about 7100~K to $\sim 0.4$ for models with $\teff\ = 8000$~K in
order to obtain a comparable value of $(F_{\rm C})_{\rm max}$ in the
\hi\ convection zone. In order to match $(F_{\rm C})_{\rm max}$ in the
\heii\ convection zone, a completely different set of $\alpha$'s (with
larger values) would be required.

As already mentioned, A-stars are excellent choices for this first
calculation since they have relatively thin surface convection zones,
so that the thermal time scales involved are not so long. In addition,
they are interesting stars in their own right, containing
high-metallicity stars (the Am stars) as well as two groups of
pulsating stars (the roAp and $\delta$ Scuti stars). In the future, it may be
possible to use the pulsating stars as probes of the subsurface
convection zones, much as has been done in the case of the Sun.

\section*{Acknowledgments}

This research was performed within project {\sl P13936-TEC} of the
Austrian Fonds zur F\"orderung der wissen\-schaft\-lichen Forschung
(FwF), and was supported by the UK Particle Physics and Astronomy
Research Council. We thank Dr.\ B.~Freytag for providing us with
results from his simulations.

\label{lastpage}




\begin{thebibliography}{99}
\bibitem{b2} Atroshchenko I.N., Gadun A.S., 1994, A\&A, 291, 635
\bibitem{b3} Biermann L., 1948, Z. Astrophys., 25, 135
\bibitem{b4} B\"ohm-Vitense E., 1958, Z. Astrophys., 46, 108
\bibitem{b5} Canuto V.M., 1992, ApJ, 392, 218
\bibitem{b6} Canuto V.M., 1993, ApJ, 416, 331
\bibitem{b7} Canuto V.M., 2000, 24th meeting of the IAU, Joint Discussion 5,
              August 2000, Manchester, England
\bibitem{b8} Canuto V.M., Minotti F., Ronchi C., Ypma R.M., Zeman O., 1994,
         J. Atm. Sci., 51 (No. 12), 1605
\bibitem{b8b} Canuto V.M., Goldman I., Mazzitelli I., 1996, ApJ, 473, 550
\bibitem{b9} Canuto V.M., Dubovikov, M.S., 1998, ApJ, 493, 834 (CD98)
\bibitem{b10} Canuto V.M., Cheng Y., Howard A., 2001, J. Atm. Sci.,
         58, 1169
\bibitem{b11} Freytag B., 1995, PhD thesis, University of Kiel
\bibitem{b12} Freytag B., Ludwig H.-G., Steffen M., 1996, A\&A, 313, 497
\bibitem{b12.5} Iglesias C.A., Rogers F.J., 1996, ApJ,  464, 943
\bibitem{b13} Kim Y.-C., Chan K.L., 1998, ApJ, 496, L121
\bibitem{b14} Kupka F., 1999a, Theory and Tests of Convection in Stellar
         Structure, A. Gimenez, E.F. Guinan and B. Montesinos,
         ASP Conf. Ser. 173, 157
\bibitem{b15} Kupka F., 1999b, ApJ, 526, L45
\bibitem{b16} Kupka F., 2001, in Proceedings of the COROT/SWG Sept.\ 2000
         meeting, edt. E. Michel, Paris
\bibitem{b17} Kupka F., 2002, ApJ, to be submitted (Paper I+III)
\bibitem{b18} Kupka F., Montgomery M.H., 2001, in Proceedings of the COROT/SWG
         Sept.\ 2000 meeting, edt. E. Michel, Paris
\bibitem{b19} Kupka F., Muthsam H.J., 2002, ApJ, to be submitted (Paper II)
\bibitem{b19a} Kurucz R.L., 1979, ApJS, 40, 1
\bibitem{b19.5} Landstreet J.D., 1998, A\&A, 338, 1041
\bibitem{b20} Muthsam H.J., G\"ob W., Kupka F., Liebich W., Z\"ochling J.,
         1995, A\&A, 293, 127
\bibitem{b21} Nordlund {\AA}, Dravins D., 1990, A\&A, 228, 155
\bibitem{b22} Pamyatnykh A.A., 1999, Acta Astronomica, 49, 119
\bibitem{b24} Renzini A., 1987, A\&A, 188, 49
\bibitem{b24.5} Rogers F.J., Swenson F.J., Iglesias C.A., 1996, ApJ,
                456, 902
\bibitem{b25} Spiegel E.A., 1957, ApJ, 126, 202
\bibitem{b26} Stein R.F., Nordlund {\AA}, 1998, ApJ, 499, 914
\bibitem{b27} Varenne O., Monier R., 1999, A\&A, 351, 247
\bibitem{b28} Xiong D.R., 1978, Chin. Astron., 2, 118
\bibitem{b29} Xiong D.R., Cheng Q.L., Deng L., 1997, ApJS, 108, 529
\end{thebibliography}
\end{document}